\documentclass[11pt]{article}
\usepackage[hmargin=0.85in,vmargin=0.8in]{geometry}
\usepackage[T1]{fontenc}

\DeclareFontFamily{T1}{lmr}{}
\DeclareFontShape{T1}{lmr}{m}{n}{<->ec-lmr10}{}
\DeclareFontShape{T1}{lmr}{m}{it}{<->ec-lmri10}{}
\DeclareFontShape{T1}{lmr}{bx}{n}{<->ec-lmbx10}{}
\usepackage{amsmath,amssymb,mathtools}
\DeclareMathAlphabet{\mathbf}{T1}{lmr}{bx}{n}
\usepackage[authoryear,round]{natbib}
\usepackage{titlesec}
\titlespacing*{\section}{0pt}{2.2ex plus .5ex minus .2ex}{1ex plus .2ex}
\titlespacing*{\subsection}{0pt}{1.8ex plus .4ex minus .2ex}{.7ex plus .2ex}
\usepackage[colorlinks=true,linkcolor=blue,citecolor=blue,urlcolor=blue]{hyperref}
\newcommand{\dd}{\mathrm{d}}
\newcommand{\lP}{\ell_P}
\newcommand{\doi}[1]{\href{https://doi.org/#1}{doi:#1}}
\title{Renormalizing the gravitational clock of a\\ causal diamond in linearized gravity}
\author{A.~M.~Silva\\[4pt]
\small Instituto de F\'isica, Universidade de S\~ao Paulo, Rua do Mat\~ao, 1371, Cidade Universit\'aria,\\
\small 05508-090, S\~ao Paulo, SP, Brazil}
\date{}
\begin{document}
\maketitle

\begin{abstract}
A recent quantization of gravity in causal diamonds restricts a gravitational clock built from exponentiated null focusing to bounded readings, and the resulting area bound underlies its finite partition function. We study the renormalization of this clock for free gravitons in Minkowski spacetime. Without subtraction, the vacuum focusing grows as $\lP^2\Lambda^4b^2$ with the frequency cutoff $\Lambda$. After normal ordering, the focusing of a single null cone is unbounded below, even after averaging over its generators, because gravitons nearly parallel to a generator keep a fixed null frequency at arbitrarily high energy. Averaging the cone's vertex over a short time window removes both problems, since the sampled focusing obeys a quantum inequality uniform in direction and radial profile. At second order in the perturbation, where the exponential clock and its linearization agree, the time-sampled clock of a diamond of radius $b$ is a self-adjoint operator bounded below by $L(b-a/b)$. Here $L<1$ is geometric and $a\propto\lP^2/\eta^4$ for sampling width $\eta b$. Its projection onto readings at most $\ell$ is nonzero whenever $Lb<\ell$, and it confines the corner area below a cap that exceeds the classical value by $8\pi a$ to first order in $a$. The clock is localized in a bounded region, so by the Reeh--Schlieder theorem its bounded-reading sector is generated from the projected vacuum by operations spacelike to the diamond. The same locality would give the full exponential clock if the vacuum had a finite exponential reading, but we show that this reading is infinite at every coupling.
\end{abstract}


\section{Introduction}

The causal-diamond quantization of \citet{Chandrasekaran} defines a diamond Hilbert space by restricting a gravitational clock to readings below a fixed length $\ell=\ell(D)$. The clock is the affine parameter along the outgoing null generators, from a cap near the past tip to the corner, and shear delays it through Raychaudhuri focusing. In the round background sector, with corner area $A=4\pi b^2$ and the cap slope at its background value, the clock takes the form
\begin{equation}\label{eq:clock}
 T_b=b\int_\epsilon^1\dd v\int\frac{\dd\Omega_n}{4\pi}\,e^{\Xi_{v,n}},
 \qquad \Xi_{v,n}=\frac12\int_{\epsilon b}^{vb}r\,\sigma_{AB}\sigma_{AB}\,\dd r,
 \qquad L=1-\epsilon ,
\end{equation}
where $v=r/b$, $0<\epsilon<1$ locates the cap, and $\sigma_{AB}$ is the shear with orthonormal screen indices. This is the linearized form of equations (3.25d), (3.26) and (4.26) of \citet{Chandrasekaran}. Classically $\Xi\ge0$, so $T_b\ge Lb$, and readings at most $\ell$ require $A\le4\pi\ell^2/L^2$. This is the area bound of equation (5.21d) in that work, which underlies its finite partition function. Quantization replaces the shear square by a composite operator, which the original construction keeps nonnegative by writing it as a sum of squares in equation (4.22) of \citet{Chandrasekaran}, so that the operator bound holds even though the ordering involves no subtraction. We ask whether a renormalized clock exists and whether the area mechanism survives renormalization.

The first of two features that shape our approach is that $\Xi$ is quadratic in the metric perturbation $h$, so that $e^\Xi=1+\Xi+O(h^4)$, while linearized gravity determines $\Xi$ only to $O(h^2)$. At $O(h^4)$ the term $\tfrac12\Xi^2$ of the exponential competes with quartic terms of the nonlinear free data and with interaction corrections, none of which the linearized theory contains, so at the order of the quantum theory we use the consistent clock is
\begin{equation}\label{eq:clock2}
 T^{(2)}_b=b\int_\epsilon^1\dd v\int\frac{\dd\Omega_n}{4\pi}\,\bigl(1+\Xi_{v,n}\bigr).
\end{equation}
It agrees with \eqref{eq:clock} to second order and has the same classical bound $T\ge Lb$. The second feature is that a quadratic composite of a free field is defined only after subtraction and smearing, and the choice of smearing decides which lower bounds hold.

For free gravitons in Minkowski spacetime we find that, without subtraction, the vacuum focusing diverges and the time-sampled clock loses all bounded readings as the cutoff is removed. Normal ordering on a single null cone does not cure this, even after averaging over generators, whereas averaging the vertex over time does (Sections~\ref{sec:obstructions} and~\ref{sec:qei}). The resulting renormalized clock is self-adjoint, caps the area at an explicit value and has a nonzero bounded-reading sector, and because it is localized in a bounded region, the Reeh--Schlieder theorem ties that sector to the vacuum (Section~\ref{sec:clock}). For the exponential response the same locality would give a clock if the vacuum had a finite reading, but that reading is infinite at every coupling (Section~\ref{sec:exp}). The exponential resummation beyond linearized gravity, curved backgrounds, the nonlinear representation of the original construction, and its state counting and entropy lie outside these statements.

\section{Optical shear of free gravitons}\label{sec:shear}

Work in four-dimensional Minkowski spacetime with $\lP^2=G\hbar$, and translate a radius-$b$ diamond in time. With vertex at $(\tau,0)$, its outgoing generators are $x^\mu(\tau,r)=(\tau+r,rn)$, $k^\mu=(1,n)$, with $n$ a unit vector and orthonormal screen vectors $E_A\perp n$. The regular-vertex Sachs solution is
\begin{equation}\label{eq:sachs}
 \sigma_{AB}(\tau,r,n)=-r^{-2}\int_0^r s^2 C^{(1)}_{AkBk}(\tau+s,sn)\,\dd s .
\end{equation}
The background shear vanishes, so this first-order shear is gauge invariant \citep{StewartWalker}.

We expand the transverse-traceless metric perturbation in plane waves,
\begin{equation}\label{eq:modes}
 h_{ij}(t,x)=\sqrt{16\pi G\hbar}\sum_{\lambda=1}^2
 \int\frac{\dd^3p}{\sqrt{2\omega(2\pi)^3}}
 \left[e^\lambda_{ij}a_\lambda(p)e^{-i\omega t+ip\cdot x}+\mathrm{h.c.}\right],
 \qquad \omega=|p| .
\end{equation}
The two polarization tensors are normalized by $e^\lambda_{ij}e^{\lambda'}_{ij}=2\delta_{\lambda\lambda'}$ and the oscillators obey $[a_\lambda(p),a^\dagger_{\lambda'}(p')]=\delta_{\lambda\lambda'}\delta^3(p-p')$, a normalization that follows from the quadratic action $(64\pi G)^{-1}\int[(\partial_t h_{ij})^2-(\nabla h_{ij})^2]\dd^4x$.

A plane wave enters the shear on the generator $n$ through its null frequency, its momentum along the screen, and a vector combining the two,
\begin{equation}\label{eq:qV}
 q=\omega-p\cdot n,\qquad p_A=p\cdot E_A,\qquad V_A=qE_A+p_An .
\end{equation}
The null frequency $q$ is small when the graviton moves nearly parallel to the generator, and the curvature component that drives the shear, $C_{AkBk}$, is proportional to $R^\lambda_{AB}(p)=\tfrac12V_A^iV_B^je^\lambda_{ij}$. Along the generator the plane wave carries the phase $e^{-i\omega\tau-iqs}$, and the Sachs integral \eqref{eq:sachs} turns $R$ into the shear amplitude
\begin{equation}\label{eq:S}
 S^\lambda_{AB}(r,p)=-r\,H(qr)\,R^\lambda_{AB}(p),\qquad H(z)=\int_0^1t^2e^{-izt}\,\dd t .
\end{equation}

The first of two properties that control everything below is that, summed over polarizations and screen indices, these amplitudes depend on the momentum only through the null frequency,
\begin{equation}\label{eq:polar}
 \sum_{\lambda,A,B}|R^\lambda_{AB}|^2=q^4,\qquad
 \sum_{\lambda,A,B}|S^\lambda_{AB}|^2=q^4r^2|H(qr)|^2\le4q^2 .
\end{equation}
The first identity follows from $(PV_A)\cdot(PV_B)=q^2\delta_{AB}$, where $P$ projects transverse to $p$. The second property is that the response function is bounded, with $|H|\le1/3$ and, by integration by parts, $|zH(z)|\le2$, so that neither the vertex nor gravitons moving along the generator produce a singularity.

The contracted vacuum covariance at fixed $(r,n)$ is $\int_0^\infty\rho_r(\omega)e^{-i\omega(\tau-\tau')}\dd\omega$, with
\begin{equation}\label{eq:density}
 \rho_r(\omega)=\frac{\lP^2}{\pi^2}\,\omega\!\int\!\dd\Omega_p\,q^4r^2|H(qr)|^2
 =\frac{2\lP^2}{\pi r^3}J(2\omega r),\qquad
 J(Z)=\int_0^Zz^4|H(z)|^2\dd z .
\end{equation}
Since $z^4|H|^2\le4z^2$ and $J(Z)=Z^3/3+O(Z^2)$, the density obeys $\rho_r(\omega)\le(64/3\pi)\lP^2\omega^3$ uniformly in $r$ and $n$, and $\rho_r(\omega)\to(16/3\pi)\lP^2\omega^3$ when $\omega r\to\infty$.

For real $g\in C^\infty_0(\mathbb R)$ with $\|g\|_2=1$ and measurable radial profiles $0\le f_v\le1$ supported in $[\epsilon b,b]$, define the time-sampled focusing
\begin{equation}\label{eq:Qdef}
 Q_{v,n}[g]=\frac12\int r f_v(r)\,\dd r\int g(\tau)^2\,\sigma_{AB}\sigma_{AB}(\tau,r,n)\,\dd\tau ,
\end{equation}
and write $:\!Q\!:$ for its normal-ordered version in the Minkowski vacuum. The sharp profile $f_v=1_{[\epsilon b,vb]}$ reproduces the cumulative focusing in \eqref{eq:clock}, and the single-cone focusing $\Xi_{v,n}(\tau_0)$ is the formal limit $g^2\to\delta(\tau-\tau_0)$. Let $\mathcal D$ denote finite-particle vectors with smooth momentum wave functions of compact support away from $p=0$.

\section{Subtraction is necessary but not sufficient}\label{sec:obstructions}

\subsection{The vacuum divergence}

The shear square in \eqref{eq:clock} multiplies two field operators at the same point, so its vacuum expectation is infinite. To see how fast it grows, we keep only gravitons with $|p|\le\Lambda$ and mark the truncated quantities by a superscript $\Lambda$ (removing modes can only lower $\rho_r$). By Wick's theorem the unsubtracted sampled focusing is then its normal-ordered part plus a constant,
\begin{equation}\label{eq:mu}
 Q^\Lambda_{v,n}[g]=\,:\!Q^\Lambda_{v,n}[g]\!:+\mu_\Lambda(v)I,\qquad
 \mu_\Lambda(v)=\frac12\int rf_v\,\dd r\int_0^\Lambda\rho_r\,\dd\omega
 \simeq\frac{2}{3\pi}\lP^2\Lambda^4\int rf_v\,\dd r .
\end{equation}
The constant is the same for every sampler $g$, because the vacuum is stationary, and it equals the vacuum expectation of the unsampled single-cone focusing. For the cumulative profile $f_v=1_{[\epsilon b,vb]}$ its leading term is $\lP^2\Lambda^4b^2(v^2-\epsilon^2)/(3\pi)$. Appendix~\ref{app:cone} shows that the corrections are of relative size $1/(\Lambda\epsilon b)$.

Because the normal-ordered part is bounded below uniformly in $\Lambda$ (Section~\ref{sec:qei}), and because $e^x\ge1+x$, this constant raises every reading of the exponential clock and of the second-order clock built from unsubtracted, time-sampled focusing, both of which read at least $Lb-La/b+b\int\mu_\Lambda\,\dd v$, with $a$ given in \eqref{eq:a}. As $\Lambda$ grows this exceeds any $\ell$, so that no state has a bounded reading, which is the fate of the sum-of-squares ordering of the original construction and the reason the vacuum term has to be subtracted.

\subsection{Normal ordering on a single cone}

Subtracting the vacuum term makes the focusing finite but removes its positivity, and on a single null cone nothing takes the place of positivity, because the normal-ordered focusing of one cone has no lower bound.

The reason is kinematic: at a fixed vertex time, a graviton's phase along the generator $n$ is $e^{-iqs}$ up to a constant, so it depends on the momentum only through the null frequency $q$, and a graviton moving almost parallel to $n$ has a small $q$ however large its energy. Labelling momenta by $(\omega,q,\varphi)$, with $\varphi$ the azimuth about $n$, the invariant measure becomes $\dd^3p/(2\omega)=\tfrac12\,\dd\omega\,\dd q\,\dd\varphi$ with $0\le q\le2\omega$. As $\omega\to\infty$ at fixed $q$ and $\varphi$, the amplitude $R^\lambda_{AB}$ tends to a nonzero limit, while by \eqref{eq:polar} the sum of its squares stays $q^4$. Gravitons of every energy therefore couple to the shear on the generator with the same strength, and smoothing along the generator does not suppress them.

To exhibit the negative values directly, take a smooth weight $f\ge0$ supported inside $(\epsilon b,b)$ and the single-cone focusing
\[
 \Xi_f(n)=\frac12\int rf(r)\,:\!\sigma_{AB}\sigma_{AB}\!:(\tau_0,r,n)\,\dd r .
\]
Superpose the vacuum with a state $\Phi_\Omega$ of two gravitons whose energies lie between $\Omega$ and $2\Omega$ and whose null frequencies are held fixed:
\[
 \psi=\cos\theta\,|0\rangle+e^{i\alpha}\sin\theta\,|\Phi_\Omega\rangle .
\]
The focusing connects the two components through its pair term, with amplitude $k_\Omega$, and acts within the two-graviton component through its number term, with value $n_\Omega\ge0$. Because gravitons of every energy couple with the same strength, both quantities grow linearly with the energy range, so that $k_\Omega$ and $n_\Omega$ are proportional to $\Omega$. Choosing $\theta$ and $\alpha$ to make the cross term negative gives
\[
 \min_{\alpha,\theta}\langle\psi,\Xi_f(n)\psi\rangle=\frac{n_\Omega}2-\Bigl(\frac{n_\Omega^2}4+|k_\Omega|^2\Bigr)^{1/2}\le-c\,\Omega ,
\]
with $c>0$ because the pair amplitude does not vanish, so the expectation decreases without bound as $\Omega$ grows.

Averaging over generators with a smooth positive weight does not help, because a graviton of energy $\Omega$ is nearly parallel only to a patch of generators of solid angle about $q/\Omega$. Dividing the sphere into about $\Omega$ such patches and placing one superposition of this kind in each, the contributions of different patches add, and the total is again of order $-\Omega$, as Appendix~\ref{app:cone} shows with explicit estimates for both constructions.

The same mechanism makes even the radially smeared shear of one cone fluctuate infinitely in the vacuum. The result is the graviton analogue of the fact that, for a free field in four dimensions, null energy averaged along a null geodesic has no lower bound \citep{FewsterRoman}, and neither has null energy averaged over a null plane \citep{SemiLocal}. In terms of angular harmonics on the cone, nearly parallel gravitons of energy $\omega$ occupy harmonics up to order $(q\omega)^{1/2}b$, so the obstruction involves arbitrarily high harmonics.

\section{Time sampling}\label{sec:qei}

Sampling in $\tau$ multiplies the pair kernel by $\widehat{g^2}(\omega+\omega')$ and the number kernel by $\widehat{g^2}(\omega-\omega')$, which damps exactly the pairs used above. In null coordinates $x^+=\tau+2r$, $x^-=\tau$, joint $(r,\tau)$ sampling spans both null directions, in line with the double-smeared null energy condition \citep{DSNEC}. The positive-square method of \citet{Fewster} and \eqref{eq:density} give, with $\widehat g(s)=\int e^{is\tau}g(\tau)\dd\tau$,
\begin{equation}\label{eq:shearqei}
 \int g^2\langle:\!\sigma_{AB}\sigma_{AB}\!:\rangle\,\dd\tau
 \ge-\frac1\pi\int_0^\infty\!\dd\alpha\int_0^\infty\!\dd\omega\,
           \rho_r(\omega)|\widehat g(\alpha+\omega)|^2
 \ge-\frac{16}{3\pi}\lP^2\|g''\|_2^2 .
\end{equation}
The last step interchanges the integrals and uses $\int_0^\nu\omega^3\dd\omega=\nu^4/4$ and $\int_0^\infty\nu^4|\widehat g(\nu)|^2\dd\nu=\pi\|g''\|_2^2$. It is a calculation for the two graviton polarizations, not a rescaled scalar inequality. Integrating \eqref{eq:shearqei} against $rf_v(r)/2$ and using $\int_{\epsilon b}^b r\,\dd r=b^2(1-\epsilon^2)/2$ gives, for every state in $\mathcal D$, every $v$, $n$ and every measurable profile $0\le f_v\le1$ supported in $[\epsilon b,b]$,
\begin{equation}\label{eq:Q}
 :\!Q_{v,n}[g]\!:\ \ge -C(b,g)\,I,\qquad
 C(b,g)=\frac{4(1-\epsilon^2)}{3\pi}\lP^2b^2\|g''\|_2^2 .
\end{equation}
The bound is uniform in $v$, $n$, $f_v$ and in the cutoff $\Lambda$, which only lowers $\rho_r$, and it survives positive normalized averaging over $v$ and $n$.

On $\mathcal D$ the sampled focusing is a symmetric operator, because the bound $|S|\le2q\le4\omega$ and the rapid decay of $\widehat{g^2}$ make its pair-creation kernel Hilbert--Schmidt, while its number kernel maps compactly supported wave functions to rapidly decaying ones, with no transverse cutoff and no radial switching derivatives.

\section{The renormalized clock}\label{sec:clock}

Fix a real, even, normalized $g_0\in C_0^\infty(\mathbb R)$ supported in $[-1,1]$ and a dimensionless $\eta>0$, and set $g_\delta(\tau)=\delta^{-1/2}g_0(\tau/\delta)$ with $\delta=\eta b$. We propose as the renormalized clock the sampled average of normal-ordered second-order clocks over vertex times,
\begin{equation}\label{eq:Tren}
 T_b=\int g_\delta(\tau)^2:\!T_b^{(2)}(\tau)\!:\,\dd\tau
 =Lb\,I+b\int_\epsilon^1\dd v\int\frac{\dd\Omega_n}{4\pi}\,:\!Q_{v,n}[g_\delta]\!:,
 \qquad f_v=1_{[\epsilon b,vb]} .
\end{equation}
Because \eqref{eq:clock2} is linear in the focusing, averaging clocks and averaging focusing agree, whereas for the exponential they would not. For classical solutions smooth on the scale $\delta$, the sampled reading differs from the single-cone reading at the center by $O(\delta^2\partial_\tau^2T)$. With $\|g_\delta''\|_2^2=\|g_0''\|_2^2/(\eta b)^4$, the bound \eqref{eq:Q} becomes $C=a/b^2$, where
\begin{equation}\label{eq:a}
 a=\frac{4(1-\epsilon^2)}{3\pi\eta^4}\,\|g_0''\|_2^2\,\lP^2 .
\end{equation}

\subsection{Existence and the area cap}

The operator \eqref{eq:Tren} is symmetric on $\mathcal D$ and obeys $T_b\ge L(b-a/b)$, so its Friedrichs extension is self-adjoint with the same lower bound. Its vacuum reading is $\langle0|T_b|0\rangle=Lb$, since normal ordering removes the vacuum term. For a coherent state of a classical solution with smooth momentum profile of compact support, $\langle T_b\rangle$ equals \eqref{eq:Tren} evaluated on that solution: a Weyl operator shifts $:\!Q\!:$ by a linear field with zero vacuum expectation plus the classical value, as Appendix~\ref{app:clock} shows together with the domain details.

Represent the round sector as a direct integral over $b$, with $A=4\pi b^2$ on each fiber, $T=\int^\oplus T_b$, and $P_\ell=1_{(-\infty,\ell]}(T)$. The function $b-a/b$ increases with $b$, so the fiber projection vanishes once $L(b-a/b)>\ell$, that is, for $b>b_{\max}=\tfrac12[\ell/L+(\ell^2/L^2+4a)^{1/2}]$. This gives the area cap
\begin{equation}\label{eq:cap}
 P_\ell AP_\ell\le A_{\max}P_\ell ,\qquad
 A_{\max}=\pi\Bigl[\frac\ell L+\Bigl(\frac{\ell^2}{L^2}+4a\Bigr)^{1/2}\Bigr]^2
 =\frac{4\pi\ell^2}{L^2}+8\pi a+O\Bigl(\frac{a^2L^2}{\ell^2}\Bigr).
\end{equation}
Conversely, if $Lb<\ell$ the fiber projection cannot annihilate the vacuum: otherwise the vacuum's spectral measure would lie in $(\ell,\infty)$, and its reading would exceed $\ell>Lb$.

The projection therefore vanishes on every fiber with $b>b_{\max}$ and is nonzero on every fiber with $b<\ell/L$. On the latter it need not be the whole fiber: states with enough focusing exceed the reading there, as they do classically. The undecided band $\ell/L\le b\le b_{\max}$ has width about $aL/\ell$, and correspondingly $A_{\max}$ exceeds the classical cap $4\pi\ell^2/L^2$ by $8\pi a$ to first order in $a$. This is an upper bound, and we do not show that the spectrum of $A$ on the range of $P_\ell$ reaches it. For the compact bump $g_0\propto\exp[-1/(1-t^2)]$ one finds $\|g_0''\|_2^2=81.408$ and hence $a=34.46\,\lP^2/\eta^4$ at $\epsilon=0.05$, a conservative coefficient that depends on the profile.

For $b<\sqrt a$ the lower bound $L(b-a/b)$ is negative, so negative readings are not excluded there. Such readings have no classical counterpart and lie outside the regime where the second-order clock is meaningful. We keep them in $P_\ell=1_{(-\infty,\ell]}(T)$, but using $1_{[0,\ell]}(T)$ instead would leave the cap \eqref{eq:cap} unchanged and change nothing for $b>\sqrt a$, where $T_b>0$.

The sampled clock reproduces the classical second-order clock in a definite sense. Coherent states give exactly the classical sampled reading, and sampling changes the focusing of radiation of frequency $\nu$ by a relative amount of order $(\eta b\nu)^2$. Quantum effects enter only through the spectrum: the lower bound $L(b-a/b)$ departs from the classical $Lb$ by $La/b$, which is small for $b\gg\sqrt a\sim\lP/\eta^2$. Within that regime, the bounds on where $P_\ell$ is supported differ from the classical ones only in the band $\ell/L\le b\le b_{\max}$. Shrinking $\eta$ improves the classical approximation and raises the Planck-scale cost as $\eta^{-4}$, so the observable must fix $\eta$ and the cap depends on that choice, a trade-off that is the explicit price of the renormalization.

In the global Fock space the field solves the wave equation everywhere, so each sampled cone is regular at its vertex and the shear at the cap is the Sachs integral over $[0,\epsilon b]$ rather than independent boundary data. This is a choice of model, since the capped phase space of the original construction allows independent cap data, and including them would require their joint covariance with the bulk field as additional input, which we do not supply. The regular-vertex model agrees with the way \citet{Chandrasekaran} treats the tip, covering the excised region by another diamond as described in footnote 3 and Fig.~10 of that work, and with the smooth-vertex case $\epsilon=0$ considered there. All constants are uniform in $\epsilon\in[0,1)$, so these results have a smooth-vertex limit with $L=1$ and $a=4\|g_0''\|_2^2\lP^2/(3\pi\eta^4)$.

\subsection{Locality and the Reeh--Schlieder theorem}

Because $g_0$ has compact support, every sampled cone of a diamond of radius $b$ lies in the compact set $K_b=\{(\tau+s,sn):|\tau|\le\eta b,\ 0\le s\le b\}$. Let $O_b$ be an open double cone containing $K_b$, and $\mathcal M(O_b)$ the von Neumann algebra generated by Weyl operators of the linearized curvature smeared in $O_b$. The renormalized clock $T_b$ is affiliated with $\mathcal M(O_b)$: all its spectral projections, in particular the fiber projection $P^{(b)}_\ell=1_{(-\infty,\ell]}(T_b)$, belong to this local algebra. The reason is that $T_b$, up to a finite constant, is an integral of squares $O_\zeta^\dagger O_\zeta$ of linear fields smeared in $K_b$, so every unitary commuting with the local algebra commutes with the clock, and Appendix~\ref{app:local} checks that this holds for the Friedrichs operator itself and not only for a formal expression.

The Reeh--Schlieder theorem \citep{ReehSchlieder} states that the vacuum is cyclic and separating for the algebra of any open region whose causal complement has nonempty interior. Applied to $\mathcal M(O_b)$, it has two consequences for the bounded-reading sector. Because the vacuum is separating, $P^{(b)}_\ell\Omega=0$ only if $P^{(b)}_\ell=0$, so whenever the sector is nonzero at all, including in the undecided band $\ell/L\le b\le b_{\max}$, the vacuum has nonzero probability of a reading at most $\ell$. Because the vacuum is also cyclic for $\mathcal M(O_b')\subset\mathcal M(O_b)'$, and $P^{(b)}_\ell$ commutes with $\mathcal M(O_b)'$,
\begin{equation}\label{eq:RS}
 P^{(b)}_\ell\mathcal F_b=\overline{\bigl\{X'P^{(b)}_\ell\Omega:\ X'\in\mathcal M(O_b)'\bigr\}}
 =\overline{\bigl\{P^{(b)}_\ell X\Omega:\ X\in\mathcal M(O_b)\bigr\}} ,
\end{equation}
where $\mathcal F_b$ is the Fock space of the fiber. Since operators in $\mathcal M(O_b)'$ commute with the clock and leave its reading unchanged, the whole bounded-reading sector is generated from the single vector $P^{(b)}_\ell\Omega$ by operations spacelike to the diamond, or equivalently spanned by local excitations of the vacuum followed by the clock projection. This is the renormalized counterpart of equation (5.20) of \citet{Chandrasekaran}, which states that boundary preparations from one state span the diamond Hilbert space. Both consequences rely on the compact time support of $g_0$, because a noncompact sampler would spread the clock over all times and leave no causal complement.

\section{Locality and the exponential clock}\label{sec:exp}

Locality offers a route to the full exponential clock \eqref{eq:clock}, now with time-sampled, normal-ordered focusing $Q_{v,n}=\,:\!Q_{v,n}[g_\delta]\!:$. Each $e^{Q_{v,n}}$ is affiliated with $\mathcal M(O_b)$, by the argument above, so every $X'\in\mathcal M(O_b)'$ commutes with it and $\|e^{Q/2}X'\Omega\|\le\|X'\|\,\|e^{Q/2}\Omega\|$. If the vacuum had a finite exponential reading, $b\int\dd v\int\dd\Omega_n\langle0|e^{Q_{v,n}}|0\rangle/4\pi<\infty$, every vector $X'\Omega$ would have a finite reading too, and since these vectors are dense by the Reeh--Schlieder theorem, the exponential clock would be densely defined. By the spectral argument of Section~\ref{sec:clock} its bounded-reading sector would then be nonzero once $\ell$ exceeds the vacuum reading, and through \eqref{eq:Q} it would obey $T_b\ge Lb\,e^{-a/b^2}$, with the corresponding area cap.

However, for every nonzero coupling, every $v>\epsilon$ and every generator, the vacuum exponential moment diverges,
\begin{equation}\label{eq:vacmoment}
 \langle0|e^{uQ_{v,n}}|0\rangle=\infty\qquad\text{for every }u>0 .
\end{equation}
The argument, given in Appendix~\ref{app:local}, again uses locality, now through the modular theory of a wedge. A finite vacuum moment would make the vacuum an analytic vector for $Q$, and because $Q$ changes particle number by at most two, $e^{sQ}\Omega$ would then have finite particle-number moments for smaller $s$. Weyl operators localized in the wedge opposite to one containing $O_b$ commute with $Q$, which bounds $e^{sQ}$ on one-particle states by the modular operator of that wedge, $\Delta=e^{2\pi K}$, where $K$ generates boosts about the wedge edge \citep{LMPR}. Compressing $e^{2sQ}$ to one-particle states and taking logarithms then forces the one-particle number kernel $A$ of $Q$ to obey $A\le c_0+c_1|K|$.

Wave packets of energy $\Lambda$ violate this bound, because the shear amplitude grows like the null frequency, so that $\langle A\rangle$ grows like $\lP^2\Lambda^2$ for them while $\langle|K|\rangle$ grows only like $\Lambda$. The same argument excludes every coherent state $W(h)\Omega$, since a Weyl shift adds only a linear field and a constant to $Q$, and the result is consistent with the heavy tails of time-averaged stress tensors of free fields in four dimensions \citep{FewsterFord}.

No reference state of these kinds has a finite exponential reading, so the Reeh--Schlieder route cannot start, and whether the exponential clock has a dense domain built from more exotic states remains open. The packets that produce the contradiction have energies far above the Planck energy for diamonds much larger than the Planck length, where linearized gravity does not apply, and this is the strongest reason to regard \eqref{eq:clock2}, rather than \eqref{eq:clock}, as the clock that linearized gravity defines.

\section{Conclusions}\label{sec:conclusions}

We have studied the gravitational clock of a causal diamond for free gravitons in Minkowski spacetime, at second order in the perturbation and in the Fock representation of the Minkowski vacuum. Without subtraction, the focusing that drives the clock has a divergent vacuum value and the clock loses all bounded readings as the cutoff is removed, while normal ordering on a single null cone fails, with or without averaging over generators, because nearly parallel gravitons couple at every energy. Averaging the cone's vertex over a short time window repairs this, since the sampled focusing then obeys a quantum inequality that is uniform in radius, direction and radial profile. The resulting clock \eqref{eq:Tren} is a self-adjoint local observable whose bounded-reading sector is nonzero, is generated from the projected vacuum by operations spacelike to the diamond in the sense of \eqref{eq:RS}, and caps the area at \eqref{eq:cap}. The exponential clock built from the same focusing, by contrast, has no finite reading in the vacuum or in any coherent state.

We therefore propose the time-sampled clock \eqref{eq:Tren} as the renormalized clock of the causal diamond. The price of this choice is explicit, because the observable refers to a family of diamonds whose vertices spread over $\delta=\eta b$, and the cap depends on that resolution through $a\propto\eta^{-4}$. Keeping a single cone with a fixed frequency cutoff is the alternative, but by Section~\ref{sec:obstructions} its lower bound degrades at least linearly with the cutoff, which makes the cutoff a physical parameter of the observable, and we do not pursue it.

A dense domain for the exponential clock, if one exists, would have to avoid the vacuum and the coherent states. In Section~5.1 of \citet{Chandrasekaran} the exponential is controlled through the truncation $\min(\widehat\Xi,\Xi_*)$ and the limit $\Xi_*\to\infty$, and for renormalized focusing that limit is exactly this open domain question. Beyond the questions treated here, curved backgrounds require locally covariant subtraction \citep{HollandsWald}, and identifying $A$, the clock and $P_\ell$ in the nonlinear representation of the original construction, together with recounting states for its finite dimension and entropy formula, are separate problems. In particular, we have not addressed whether the larger cap affects the logarithmic coefficient of the entropy.

In $2+1$ dimensions gravity has no local degrees of freedom, the shear vanishes identically, and the clock depends only on how the corner length is distributed and on the frame at the vertex, so none of the renormalization questions studied here arise. A companion paper in preparation applies the same construction to $(2+1)$-dimensional gravity, where its corner sector can be compared directly with the exact quantization of causal diamonds by \citet{AndradeJacobson}, and where the contribution of the corner modes can be isolated from that of gravitons.

Within these limits the answer to the opening question is positive, since the divergence concerns the clock as written rather than the program, and once its vertex is sampled in time the clock is renormalizable at second order and is a local observable to which the Reeh--Schlieder theorem applies. The area mechanism behind the finite partition function survives, with a proved cap that exceeds the classical one by a Planck-scale amount fixed by the sampling resolution.

\appendix

\section{Single-cone focusing}\label{app:cone}

\subsection*{Conventions}
Let $0\le\epsilon<1$ and $L=1-\epsilon$. For a generator $n$ with screen basis $E_A$, a momentum $p$ has energy $\omega=|p|$, null frequency $q=\omega-p\cdot n\in[0,2\omega]$ and azimuth $\varphi$ about $n$. With $\cos\Theta_p=\hat p\cdot n$ one has $q=\omega(1-\cos\Theta_p)$, $\dd^3p=\omega\,\dd\omega\,\dd q\,\dd\varphi$ and $\dd^3p/(2\omega)=\tfrac12\dd\omega\,\dd q\,\dd\varphi$. From \eqref{eq:modes}, \eqref{eq:sachs} and \eqref{eq:S},
\begin{equation}\label{eq:sigmaexp}
 \sigma_{AB}(\tau,r,n)=\frac{\lP}{\pi}\sum_\lambda\int\frac{\dd^3p}{\sqrt\omega}
 \bigl[S^\lambda_{AB}(r,p)e^{-i\omega\tau}a_\lambda(p)+\mathrm{h.c.}\bigr].
\end{equation}
At fixed direction $\hat p$, both $q$ and $V_A$ are proportional to $\omega$, so $R^\lambda_{AB}=\omega^2\widehat R^\lambda_{AB}(\hat p)$, and $R$ is real. For $z>0$,
\begin{equation}\label{eq:Hsplit}
 H(z)=e^{-iz}h_1(z)+\frac{2i}{z^3},\qquad h_1(z)=\frac iz+\frac2{z^2}-\frac{2i}{z^3},\qquad |h_1^{(k)}(z)|\le c_kz^{-1-k}\ \ (z\ge1),
\end{equation}
and in particular $1-2/z-4/z^2\le|zH(z)|\le1+2/z+4/z^2$.

\subsection*{The coincidence divergence}
Using $(1-x)^2\ge1-2x$ and the bounds after \eqref{eq:Hsplit}, for $Z\ge4$
\[
 \frac{Z^3}3-2Z^2-8Z\le J(Z)\le\frac{Z^3}3+2Z^2+12Z+16\log Z+16,
\]
where the upper bound also uses $|H|\le1/3$ on $[0,1]$. Since $\int_0^\Lambda\rho_r\dd\omega=(\lP^2/\pi r^4)\int_0^{2\Lambda r}J(Z)\dd Z$, this gives $\int_0^\Lambda\rho_r\dd\omega=(4/3\pi)\lP^2\Lambda^4[1+O((\Lambda r)^{-1})]$, with relative remainder at most $4/(\Lambda r)$ plus lower powers, so \eqref{eq:mu} follows for $r\ge\epsilon b$. The cutoff field is a free field on the modes $|p|\le\Lambda$, so its symmetric square is its normal-ordered square plus the coincident vacuum two-point function. That function is independent of $\tau$ and integrates against $g^2$ to the same constant because $\|g\|_2=1$.

For the clock statements, $e^Q\ge1+Q$ by functional calculus for each $(v,n)$. The bound \eqref{eq:Q} holds for the cutoff field with the same constant, so $Q^\Lambda_{v,n}\ge\mu_\Lambda(v)-a/b^2$ for the sampling of Section~\ref{sec:clock}. Integrating over $(v,n)$ bounds both clocks below by $Lb-La/b+b\int\mu_\Lambda\dd v$. A self-adjoint $T\ge c$ has $1_{(-\infty,\ell]}(T)=0$ if $c>\ell$, which happens once $b\int\mu_\Lambda\dd v>\ell+La/b-Lb$.

\subsection*{The limiting single-generator amplitude}
Write $\hat p=\cos\Theta\,n+\sin\Theta\,t$ with $t=\cos\varphi E_1+\sin\varphi E_2$, and $t'=\cos\Theta\,t-\sin\Theta\,n$. Since $e^\lambda_{ij}\hat p^j=0$, one has $e^\lambda(E_A,n)=-\sin\Theta\,e^\lambda(E_A,t')$ and $e^\lambda(n,n)=\sin^2\Theta\,e^\lambda(t',t')$. With $p_A=\omega\sin\Theta\,t_A$ and $\omega\sin^2\Theta=2q-q^2/\omega$, each term of
\[
 2R^\lambda_{AB}=q^2e^\lambda_{AB}+qp_Be^\lambda(E_A,n)+qp_Ae^\lambda(n,E_B)+p_Ap_Be^\lambda(n,n)
\]
is a polynomial in $q$, $\sin\Theta$ and $\omega\sin^2\Theta$ times components of $e^\lambda$ in the frame $(E_A,t')$. Choosing polarization tensors continuous in $\hat p$ near $n$, and using $\sin\Theta=(2q/\omega)^{1/2}[1+O(q/\omega)]$, $R^\lambda_{AB}$ converges as $\omega\to\infty$ at fixed $(q,\varphi)$ to a limit $R^\lambda_{\infty,AB}(q,\varphi)$, with error $O((q/\omega)^{1/2})$ uniformly on compact sets.

\subsection*{One generator}
Fix the generator $n$ and, using time-translation invariance, set $\tau_0=0$. One-graviton states are labelled by $X=(\omega,q,\varphi,\lambda)$, with measure $\dd X=\dd\omega\,\dd q\,\dd\varphi$ summed over the polarization $\lambda$. With the rescaled oscillators $b(X)=\omega^{1/2}a_\lambda(p)$, which obey $[b(X),b^\dagger(X')]=\delta(X-X')$, equation \eqref{eq:sigmaexp} becomes $\sigma_{AB}=(\lP/\pi)\int\dd X\,[S_{AB}(r,X)b(X)+\mathrm{h.c.}]$. The single-cone focusing then splits into a pair part and a number part,
\begin{equation}\label{eq:kernels}
 \Xi_f(n)=\frac{\lP^2}{2\pi^2}\iint\bigl[K\,b(X)b(X')+\mathrm{h.c.}\bigr]\dd X\,\dd X'
 +\frac{\lP^2}{\pi^2}\iint N\,b^\dagger(X)b(X')\,\dd X\,\dd X',
\end{equation}
with kernels
\[
 K(X,X')=\int rf\,S_{AB}(r,X)S_{AB}(r,X')\,\dd r,\qquad
 N(X,X')=\int rf\,\overline{S_{AB}(r,X)}S_{AB}(r,X')\,\dd r .
\]
By the previous subsection, both kernels approach limits $K_\infty$ and $N_\infty$ that depend only on $(q,\varphi,\lambda)$ and $(q',\varphi',\lambda')$ once both energies are large. This independence of the energy is what makes the focusing unbounded below.

To exploit it we use two gravitons whose energies are of order $\Omega$ and whose null frequencies and azimuths have a fixed profile. Choose a real normalized energy profile $\chi\in C_0^\infty((1,2))$ and a real normalized profile $u$ in $(q,\varphi)$, smooth and with $q>0$ on its support. The rescaled profile $\chi_\Omega(\omega)=\Omega^{-1/2}\chi(\omega/\Omega)$ is still normalized and is supported between $\Omega$ and $2\Omega$. Let $\Phi_\Omega$ be the state of two gravitons of equal polarization with wave function
\[
 \Phi_\Omega(X,X')=2^{-1/2}\,\delta_{\lambda\lambda'}\,\chi_\Omega(\omega)\chi_\Omega(\omega')\,u(q,\varphi)\,u(q',\varphi') .
\]
Only the pair part of \eqref{eq:kernels} connects $\Phi_\Omega$ with the vacuum, and only the number part acts within the two-graviton sector. The two matrix elements that matter are therefore
\begin{gather*}
 k_\Omega=\langle0,\Xi_f\Phi_\Omega\rangle=\frac{\lP^2}{2\pi^2}\sum_\lambda\iint K^{\lambda\lambda}\,\chi_\Omega\chi_\Omega\,uu\,\dd X\,\dd X',\\
 n_\Omega=\langle\Phi_\Omega,\Xi_f\Phi_\Omega\rangle=\frac{\lP^2}{\pi^2}\sum_\lambda\iint N^{\lambda\lambda}\,\chi_\Omega\chi_\Omega\,uu\,\dd X\,\dd X',
\end{gather*}
where the profiles are evaluated at $X$ and $X'$. For large $\Omega$ the kernels may be replaced by their limits, and the energy integrals then factor out, each contributing $\int\chi_\Omega\,\dd\omega=\Omega^{1/2}\int\chi$. This gives
\[
 k_\Omega=\frac{\lP^2}{2\pi^2}\,\Omega\Bigl(\int\chi\Bigr)^2\kappa_u+o(\Omega),\qquad
 n_\Omega=\frac{\lP^2}{\pi^2}\,\Omega\Bigl(\int\chi\Bigr)^2\nu_u+o(\Omega),
\]
with $\kappa_u=\sum_\lambda\iint K^{\lambda\lambda}_\infty uu$ and $\nu_u=\sum_\lambda\iint uN^{\lambda\lambda}_\infty u\ge0$. The remainders are $o(\Omega)$ because the kernels converge uniformly on the support of the profiles and the double integral of $|\chi_\Omega(\omega)\chi_\Omega(\omega')|$ equals $\Omega(\int\chi)^2$.

For the superposition $\psi=\cos\theta\,|0\rangle+e^{i\alpha}\sin\theta\,|\Phi_\Omega\rangle$ the expectation is $n_\Omega\sin^2\theta+2\sin\theta\cos\theta\,\mathrm{Re}(e^{i\alpha}k_\Omega)$. Choosing $\alpha$ so that the last term equals $-2|k_\Omega|\sin\theta\cos\theta$ and then minimizing over $\theta$ gives
\[
 \min_{\alpha,\theta}\langle\psi,\Xi_f\psi\rangle=\frac{n_\Omega}2-\Bigl(\frac{n_\Omega^2}4+|k_\Omega|^2\Bigr)^{1/2}
 =-\Omega\,\frac{\lP^2}{\pi^2}\Bigl(\int\chi\Bigr)^2\Bigl[\Bigl(\frac{\nu_u^2}4+\frac{|\kappa_u|^2}4\Bigr)^{1/2}-\frac{\nu_u}2\Bigr]+o(\Omega),
\]
which is at most $-c\,\Omega$ with $c>0$ as soon as $\kappa_u\neq0$.

A profile with $\kappa_u\neq0$ exists because the limiting pair kernel does not vanish on its diagonal. Summed over polarizations, the diagonal is $\sum_\lambda K^{\lambda\lambda}_\infty(X,X)=q^4G(q)$ with $G(q)=\int r^3fH(qr)^2\dd r$, and $G(0)=\tfrac19\int r^3f\,\dd r>0$. For a nonnegative $u$ concentrated near a point with small $q>0$, continuity of $K_\infty$ then gives $\kappa_u\simeq q^4G(q)(\int u)^2\neq0$.

\subsection*{An average over generators}
Now let $w\ge0$ be a continuous weight on the sphere with $\int w\,\dd\Omega_n=1$, take $f\in C_0^\infty((r_1,r_2))$ with $r_1>0$, and consider $\Xi_w=\int w(n)\,\Xi_f(n)\,\dd\Omega_n$. A graviton of energy $\Omega$ is nearly parallel to a generator only when their directions differ by an angle of order $\Omega^{-1/2}$. A single superposition of the kind used above therefore couples to a solid angle of generators of order $\Omega^{-1}$, and its contribution to $\Xi_w$ is of order one rather than of order $\Omega$. We compensate by using about $\Omega$ such superpositions at once, placed in disjoint cells of momentum space.

For large $\Omega$, choose $M_\Omega\ge c_w\Omega$ directions $n_c$ in the region where $w\ge\tfrac12\max w$, with pairwise angular separation at least $3\zeta_*\Omega^{-1/2}$ for a fixed $\zeta_*>0$. The cell around $n_c$ contains the momenta with energy between $\Omega$ and $2\Omega$ and direction within an angle $\zeta_*\Omega^{-1/2}$ of $n_c$, so different cells are disjoint. Near $n_c$ we use stretched coordinates, writing a momentum as $\omega=\Omega s$, $\hat p=\exp_{n_c}(\zeta\Omega^{-1/2})$ and a generator as $n=\exp_{n_c}(\xi\Omega^{-1/2})$, with $\zeta,\xi\in\mathbb R^2$. In these coordinates $\dd^3p=\Omega^2s^2\dd s\,\dd^2\zeta$, $\dd\Omega_n=\Omega^{-1}\dd^2\xi\,[1+O(|\xi|^2/\Omega)]$ and $q=\tfrac12s|\zeta-\xi|^2[1+O(|\zeta-\xi|^2/\Omega)]$, so the null frequency stays of order one for generators within the angular range of the cell. In each cell we take the one-graviton profile $\phi_c(p)=\Omega^{-1}F(s,\zeta)$, where $F$ is smooth, supported in $1<s<2$ and $|\zeta|<\zeta_*$, normalized by $\int s^2|F|^2\dd s\,\dd^2\zeta=1$, and the same in every cell up to rotation. With $\Phi_c$ the two-graviton state of wave function $2^{-1/2}\delta_{\lambda\lambda'}\phi_c(p)\phi_c(p')$, the trial state is
\[
 \psi=\bigotimes_c\bigl[\cos\theta\,|0\rangle_c+e^{i\alpha}\sin\theta\,|\Phi_c\rangle\bigr].
\]

The expectation of $\Xi_w$ in this state is a sum of independent cell contributions. Each factor of $\psi$ contains either no graviton or two gravitons from its own cell, and the cells are disjoint, so every expectation $\langle\psi,a^\#(p)a^\#(p')\psi\rangle$ with $p$ and $p'$ in different cells vanishes, as does every expectation involving momenta outside all cells. Each cell contribution has the form found for one generator, $n_c\sin^2\theta+2\sin\theta\cos\theta\,\mathrm{Re}(e^{i\alpha}k_c)$, with
\[
 k_c=\frac{\lP^2}{2\pi^2}\sum_\lambda\!\int\!\dd\Omega_n\,w(n)\!\int\! rf\,\Psi^\lambda_{n,AB}\Psi^\lambda_{n,AB}\,\dd r,\qquad
 n_c=\frac{\lP^2}{\pi^2}\sum_\lambda\!\int\!\dd\Omega_n\,w(n)\!\int\! rf\,|\Psi^\lambda_{n,AB}|^2\,\dd r .
\]
Here $\Psi^\lambda_{n,AB}(r)=\int\dd^3p\,\omega^{-1/2}\phi_c(p)S^\lambda_{AB}(r,p,n)$ is the shear profile that one graviton of the cell produces on the generator $n$. Because $|\Psi\cdot\Psi|\le|\Psi|^2$, the contribution of any set of generators to $|k_c|$ is at most half of its contribution to $n_c$.

The cell contributions approach limits that do not depend on $\Omega$. In stretched coordinates each wave function contributes $\dd^3p\,\omega^{-1/2}\phi_c=\Omega^{1/2}s^{3/2}F\,\dd s\,\dd^2\zeta$ and the generator measure contributes $\Omega^{-1}$, so the powers of $\Omega$ cancel. As $\Omega\to\infty$,
\begin{equation}\label{eq:celllimit}
 k_c\to w(n_c)\,\frac{\lP^2}{2\pi^2}\,\kappa_F,\qquad n_c\to w(n_c)\,\frac{\lP^2}{\pi^2}\,\nu_F,
\end{equation}
uniformly in $c$, where $\kappa_F$ and $\nu_F$ are the same expressions evaluated with the limiting amplitudes $R^\lambda_{\infty,AB}$ found above, in stretched coordinates and with $\xi$ ranging over all of $\mathbb R^2$. Given \eqref{eq:celllimit}, each cell contributes at most $-\tfrac14w_{\max}m_F$ for large $\Omega$, where $m_F=(\lP^2/\pi^2)[(\nu_F^2/4+|\kappa_F|^2/4)^{1/2}-\nu_F/2]$ is positive whenever $\kappa_F\neq0$, and the total is at most $-\tfrac14c_ww_{\max}m_F\,\Omega$.

To establish \eqref{eq:celllimit} we separate the generators near the cell, $|\xi|\le\Theta$, from those far from it. Near the cell the stretched integrands are bounded and converge uniformly as $\Omega\to\infty$, uniformly in $c$ by rotation covariance and the uniform continuity of $w$, so they converge to the corresponding parts of $\kappa_F$ and $\nu_F$. Far from the cell, with $\theta$ the angle between $n$ and $n_c$, we show below that
\begin{equation}\label{eq:tail}
 \int_{\theta\ge\Theta\Omega^{-1/2}}\dd\Omega_n\int rf\sum_{\lambda,A,B}|\Psi^\lambda_{n,AB}(r)|^2\dd r\le\frac{C_F}{\Theta^2},\qquad
 \Theta\ge\Theta_F=\max\bigl\{2\zeta_*,\pi(2/r_1)^{1/2}\bigr\},
\end{equation}
with $C_F$ independent of $\Omega$ and of the cell. This bound controls the far part of $n_c$ and hence of $k_c$, and it holds equally for the limiting integrals, so letting $\Theta\to\infty$ gives \eqref{eq:celllimit}.

The bound \eqref{eq:tail} expresses the weak shear that a cell's gravitons produce on generators they do not travel along. For $\theta\ge\Theta\Omega^{-1/2}$ every direction in the cell makes an angle of at least $\theta/2$ with $n$, so $q\ge\Omega\theta^2/(2\pi^2)$ and $qr\ge1$ on the support of $f$. We split the response function as in \eqref{eq:Hsplit} into the nonoscillatory term $2i/z^3$ and the oscillatory term $e^{-iz}h_1(z)$, and treat the two parts of the amplitude separately. The nonoscillatory part obeys $|S^{\rm no}|=|2R/(q^3r^2)|\le2/(qr^2)$, and together with $\|\phi_c\|_1\le c\,\Omega\|\phi_c\|_2$ this gives $|\Psi^{\rm no}|^2\le C/(\Omega\theta^4r^4)$. Along each direction $\hat p$ the oscillatory part has amplitude $\omega^{7/2}\phi_c\,\widehat R\,h_1(qr)$, and we integrate it by parts twice in $\omega$ using $\partial_\omega e^{-iqr}=-ir(1-\cos\Theta_p)e^{-iqr}$. Each derivative with respect to $\omega$ costs at most a factor $C/\Omega$, because $\partial_\omega^k\phi_c=\Omega^{-1-k}\partial_s^kF$ and $|\partial_\omega^k h_1(qr)|\le c_k(qr)^{-1}\omega^{-k}$. With $|\widehat R|\le(1-\cos\Theta_p)^2$ and a cell solid angle of order $\Omega^{-1}$, this gives $|\Psi^{\rm osc}|\le C\Omega^{-1/2}r^{-2}\theta^{-2}$. Both parts together satisfy $\sum|\Psi|^2\le C/(\Omega\theta^4r^4)$, and integrating $r^{-3}f$ over $r$ and $(\Omega\theta^4)^{-1}$ over $\theta\ge\Theta\Omega^{-1/2}$, with $\dd\Omega_n\le2\pi\theta\,\dd\theta$, gives \eqref{eq:tail}.

Finally, $\kappa_F\neq0$ for a suitable $F$, because the generator-averaged pair kernel is continuous and does not vanish on its diagonal. Let $K_\infty(X,X',\xi)$ denote the limiting pair kernel for a generator at stretched position $\xi$, so that $\kappa_F=\sum_\lambda\iint\tilde F\tilde F\int\dd^2\xi\,K^{\lambda\lambda}_\infty$ with $\tilde F=s^{3/2}F$. For large $|\xi|$ both null frequencies grow like $|\xi|^2$. In $K_\infty=\int r^3fH(qr)H(q'r)\dd r\,R_\infty\cdot R'_\infty$ the terms containing $e^{-iqr}$ or $e^{-iq'r}$ then decay faster than any power after integration by parts in $r$, while the remaining term is $-4(qq')^{-3}\int fr^{-3}\dd r\,R_\infty\cdot R'_\infty=O((qq')^{-1})$. Hence $|K_\infty|\le C(1+|\xi|)^{-4}$ uniformly for $X,X'$ in compact sets, and the averaged kernel $\tilde K(X,X')=\int\dd^2\xi\,K_\infty$ is continuous. Its diagonal can be computed exactly, since $\sum_\lambda K^{\lambda\lambda}_\infty=q^4G(q)$ there and $\dd^2\xi=s^{-1}\dd q\,\dd\varphi$:
\[
 \sum_\lambda\tilde K^{\lambda\lambda}(X,X)=\frac{2\pi}s\int_0^\infty q^4G(q)\,\dd q=\frac{2\pi}s\,I\int\frac{f}{r^2}\dd r,\qquad
 I=-24i\int_0^1\!\!\int_0^1\frac{t^2s'^2}{(t+s')^5}\dd t\,\dd s'=-\frac{5i}4 .
\]
Here $I=\int_0^\infty z^4H(z)^2\dd z$ is evaluated with an Abel factor, using $H^2=\iint t^2s'^2e^{-iz(t+s')}$ and $\int_0^\infty z^4e^{-izx}\dd z=24/(ix)^5$. The integral over $q$ converges, since $q^4G(q)=-4q^{-2}\int fr^{-3}\dd r$ plus rapidly decaying terms, and the smoothing in $r$ justifies the interchange of integrals. Because the diagonal is nonzero, a nonnegative $F$ concentrated near one point gives $\kappa_F\neq0$ by continuity.

\section{The renormalized clock}\label{app:clock}

\subsection*{The time-sampled bound}
For fixed $(r,n)$ and each $A,B$, let $O_\alpha=\int g(\tau)e^{i\alpha\tau}\sigma_{AB}(\tau,r,n)\dd\tau$. Fourier inversion and the symmetry of the normal-ordered product give, as quadratic forms on $\mathcal D$,
\[
 \int g^2:\!\sigma_{AB}\sigma_{AB}\!:\dd\tau=\frac1\pi\int_0^\infty:\!O_\alpha^\dagger O_\alpha\!:\dd\alpha
 \ \ge-\frac1\pi\int_0^\infty\dd\alpha\int_0^\infty\dd\omega\,\rho_r(\omega)|\widehat g(\alpha+\omega)|^2 ,
\]
since $O_\alpha^\dagger O_\alpha\ge0$ and the subtracted vacuum expectation is the right side \citep{Fewster}. With $\rho_r\le(64/3\pi)\lP^2\omega^3$ and $\nu=\alpha+\omega$ this gives \eqref{eq:shearqei}, and integrating against $rf_v/2$ gives \eqref{eq:Q}, which also holds for the cutoff field because its density $\rho_r1_{\omega\le\Lambda}$ is smaller.

\subsection*{Kernel estimates}
The pair-creation kernel of the sampled focusing is, up to polarization labels,
\[
 P(p,p')=\frac{\lP^2}{2\pi^2}\,\frac{\widehat{g^2}(-\omega-\omega')}{\sqrt{\omega\omega'}}\int rf_v\,S_{AB}(r,p)S_{AB}(r,p')\,\dd r .
\]
Since $\sum|S|^2\le4q^2\le16\omega^2$ and $\int rf_v\le b^2/2$,
\[
 |P(p,p')|\le\frac{4\lP^2b^2}{\pi^2}\,(\omega\omega')^{1/2}\,\bigl|\widehat{g^2}(\omega+\omega')\bigr| ,
\]
so $P$ is Hilbert--Schmidt because $\widehat{g^2}$ is a Schwartz function. The bound is uniform in $(v,n)$ and survives averaging by Minkowski's integral inequality. The number kernel carries $\widehat{g^2}(\omega-\omega')$ and maps wave functions supported in $\omega'\le\Omega_0$ to functions decaying faster than any power of $\omega$. The sampled focusing therefore maps $\mathcal D$ into Fock space and is symmetric there.

\subsection*{Self-adjointness, coherent states and the direct integral}
For $f_v=1_{[\epsilon b,vb]}$, $\tfrac12\int rf_v\dd r\le\tfrac14b^2(1-\epsilon^2)$ and $\|g_\delta''\|_2^2=\|g_0''\|_2^2/(\eta b)^4$, so each $:\!Q_{v,n}\!:\,\ge-a/b^2$, and integrating over a set of measure $L$ gives $T_b\ge Lb-La/b$ on $\mathcal D$. By the kernel estimates, $T_b$ is densely defined, symmetric and bounded below. Its Friedrichs extension is self-adjoint with the same bound, and its form domain is the closure of $\mathcal D$ in the form norm.

For $h$ smooth with compact support $K$ away from $p=0$, $W(h)^*a(p)W(h)=a(p)+h(p)$ gives $W(h)^*:\!Q\!:W(h)=\,:\!Q\!:+Q_{\rm lin}[h]+Q_{\rm cl}[h]$, with $Q_{\rm lin}$ linear in the field and $Q_{\rm cl}[h]$ the classical sampled focusing. Let $\psi_m\in\mathcal D$ be the normalized truncation of $W(h)|0\rangle$ to at most $m$ gravitons. On vectors with momenta in $K$, each term of $T_b+La/b$ is bounded by a constant times $\langle(\mathcal N+1)^2\rangle$, with $\mathcal N$ the number operator, because the kernels restricted to $K\times K$ are bounded and the pair terms are Hilbert--Schmidt. Since $(\mathcal N+1)\psi_m$ converges, $\psi_m$ is Cauchy in form norm, and $W(h)|0\rangle$ lies in the form domain with the limiting expectation.

For $\psi\in\mathcal D$, $b\mapsto\langle\psi,T_b\psi\rangle$ is continuous, because $\epsilon b$, $vb$ and $\delta=\eta b$ enter continuously. Let $\mathcal D_*\subset\mathcal D$ be countable and dense in each form norm. The variational formula $\langle\chi,(T_b+c)^{-1}\chi\rangle=\sup_{u\in\mathcal D_*}\{2\operatorname{Re}\langle\chi,u\rangle-\langle u,(T_b+c)u\rangle\}$, valid for $c>La/b-Lb+1$ locally uniformly in $b$, makes the resolvents measurable. The direct integral $T=\int^\oplus T_b\,\dd\nu(b)$ is then self-adjoint and commutes strongly with $A$. Functional calculus acts fiberwise, so $P_\ell=\int^\oplus1_{(-\infty,\ell]}(T_b)$, which vanishes on fibers with $L(b-a/b)>\ell$ and so gives \eqref{eq:cap}. A direct-integral projection is nonzero when $\nu$ gives positive weight to radii with $Lb<\ell$, although no statement about finite rank follows.

\section{Locality and the vacuum exponential reading}\label{app:local}

\subsection*{Locality of the clock}
Fix $b$ and write $\zeta=(v,n,\alpha,r,i)$ with $\alpha\ge0$ and $i$ a screen component. Let $O_\zeta=\int g_\delta(\tau)e^{i\alpha\tau}\sigma_i(\tau,r,n)\dd\tau$. Through the Sachs integral \eqref{eq:sachs}, its real and imaginary parts are the linearized curvature smeared against real distributions supported in $K_b$. Their momentum coefficients contain $\widehat{g_\delta}(\alpha\mp\omega)$ times amplitudes of polynomial growth, so they have finite one-particle norm. Mollifying these smearings within $O_b$ approximates them in one-particle norm, so the corresponding Weyl operators lie in $\mathcal M(O_b)$. The closed fields $O_\zeta=a(u_\zeta)+a^\dagger(w_\zeta)$ are therefore affiliated with $\mathcal M(O_b)$.

Integrating the positive-square identity of Appendix~\ref{app:clock} over $(v,n)$ gives, for $\psi\in\mathcal D$,
\begin{equation}\label{eq:possq}
 \langle\psi,(T_b-Lb+bC)\psi\rangle=b\int\|O_\zeta\psi\|^2\dd\mu(\zeta),\qquad C=\int\|w_\zeta\|^2\dd\mu<\infty,
\end{equation}
where $\dd\mu$ collects $\dd v\,\dd\Omega_n/4\pi$, $rf_v(r)\dd r/2$, $\dd\alpha/\pi$ and the screen sum. The constant $C$ is the vacuum subtraction of \eqref{eq:shearqei}, finite and uniform in $(v,n)$. The right side defines a closed positive form on its maximal domain. A unitary $U\in\mathcal M(O_b)'$ commutes with every $O_\zeta$, so it preserves that domain and the form, and the associated operator is affiliated with $\mathcal M(O_b)$. It coincides with the Friedrichs extension of $T_b-Lb+bC$ because $\mathcal D$ is a core for the maximal form. Damping with $e^{-tN}$ and truncating particle number approximate any vector of the maximal domain in the graph norm of $\psi\mapsto(O_\zeta\psi)_\zeta$, the creation parts being controlled by the finite constant $C$ and the annihilation parts by $|S|\le4\omega$ and the rapid decay of $\widehat{g_\delta}$, which make compact energy support a core for the one-particle analysis map. Hence $T_b$ and all its spectral projections are affiliated with $\mathcal M(O_b)$. The same argument applies to each $:\!Q_{v,n}\!:$ separately, and therefore to each $e^{Q_{v,n}}$.

\subsection*{The vacuum exponential reading}
Fix $(v,n)$ with $v>\epsilon$ and write $Q=\,:\!Q_{v,n}[g_\delta]\!:$, with $Q\ge-c$ by \eqref{eq:Q}. Its one-particle compression is the number kernel $A$, since the pair terms change particle number by two. We suppose that $\Omega\in D(e^{tQ})$ for some $t>0$ and derive a contradiction in four steps.

The first step obtains particle-number moments from analyticity: since $Q\ge-c$, $\Omega\in D(e^{t|Q|})$ and $\|Q^n\Omega\|\le n!\,t^{-n}\|e^{t|Q|}\Omega\|$. The operator $Q$ maps a vector with at most $m$ gravitons to one with at most $m+2$, as one sees by testing against smooth vectors in higher particle sectors. Hence $Q^n\Omega$ has at most $2n$ gravitons, and the series for $\Psi_s=e^{sQ}\Omega$ converges in every norm $\|(N+1)^k\cdot\|$ for $0<s<t$. In particular $d_s^2=\|(N+1)^{1/2}\Psi_s\|^2<\infty$.

For the second step, which uses wedge locality, let $W$ be a wedge containing $O_b$, $W'$ its causal complement, and $\mathcal K$ the closed real subspace of one-particle vectors localized in $W'$. Its Tomita operator $S(h+ik)=h-ik$ has $S^*S=\Delta=e^{2\pi K}$, with $K$ the boost generator about the edge of $W$ in a suitable sign convention \citep{LMPR}. For $h\in\mathcal K$ the Weyl operator $e^{i\lambda\Phi(h)}$ lies in $\mathcal M(W')\subset\mathcal M(O_b)'$ and commutes with $e^{sQ}$. Differentiating $e^{sQ}e^{i\lambda\Phi(h)}\Omega=e^{i\lambda\Phi(h)}\Psi_s$ at $\lambda=0$ gives $e^{sQ}\Phi(h)\Omega=\Phi(h)\Psi_s$. For $u=h+ik$ with $h,k\in\mathcal K$ this reads $e^{sQ}a^\dagger(u)\Omega=[a(Su)+a^\dagger(u)]\Psi_s$, so
\[
 \|e^{sQ}a^\dagger(u)\Omega\|^2\le2d_s^2\bigl(\|u\|^2+\|\Delta^{1/2}u\|^2\bigr).
\]

The third step is a logarithmic compression, in which $J:u\mapsto a^\dagger(u)\Omega$ denotes the one-particle embedding, $X=e^{2s(Q+c)}\ge1$, and $T$ the operator of the form $u\mapsto\|X^{1/2}Ju\|^2$. The previous bound gives $1\le T\le C_s(1+\Delta)$ with $C_s=2e^{2sc}d_s^2$. The compression inequality $J^*(X+\lambda)^{-1}J\ge(T+\lambda)^{-1}$, integrated in $\log X=\int_0^\infty[(1+\lambda)^{-1}-(X+\lambda)^{-1}]\dd\lambda$, gives $J^*(\log X)J\le\log T$. Since $J^*(Q+c)J=A+c$ and the logarithm is operator monotone,
\[
 2s(A+c)\le\log T\le\log(2C_s)+2\pi|K|,\qquad\text{so}\qquad A\le c_0+\frac\pi s|K| .
\]

In the last step we take normalized one-graviton packets $u_\Lambda$ of fixed width, centred at energy $\Lambda$ and null frequency $\gamma\Lambda$, $0<\gamma<2$, with respect to the boost axis $n$. On the support of $f_v$, $S_0(r,q)=-rq^2H(qr)=-iq\,e^{-iqr}+O(1)$, which gives $\langle u_\Lambda,Au_\Lambda\rangle\ge c_*\Lambda^2+o(\Lambda^2)$ with $c_*>0$ proportional to $\lP^2$. In the variables $(\omega,q)$ the boost acts as the first-order operator $i[(\omega-q)\partial_\omega-q\partial_q]$, and translating the wedge adds energy and momentum, so $\langle u_\Lambda,|K|u_\Lambda\rangle\le\|Ku_\Lambda\|=O(\Lambda)$. For large $\Lambda$ this contradicts the bound on $A$, so $\Omega\notin D(e^{tQ})$ for every $t>0$, which is \eqref{eq:vacmoment}.

For a coherent state $W(h)\Omega$, the operator $W(h)^*QW(h)$ differs from $Q$ by a linear field and a constant. It is affiliated with the same algebra, since $W(h)$ normalizes every local Weyl algebra, and its one-particle compression is $A$ plus a constant. Repeating the four steps shows $W(h)\Omega\notin D(e^{tQ})$ for every $t>0$.

\subsection*{CRediT authorship contribution statement}
A.~M.~Silva: Conceptualization, Methodology, Software, Formal analysis, Validation, Writing -- original draft, Writing -- review \& editing.

\subsection*{Declaration of competing interest}
The author declares that they have no known competing financial interests or personal relationships that could have appeared to influence the work reported in this paper.

\subsection*{Funding}
This research did not receive any specific grant from funding agencies in the public, commercial, or not-for-profit sectors.

\subsection*{Declaration of generative AI and AI-assisted technologies in the writing process}
During preparation, Claude (Anthropic) assisted with derivations, calculations, literature searches and drafting. OpenAI Codex assisted with the revision, mathematical derivations, numerical checks, reference verification and proofreading. The author is responsible for reviewing and validating the analysis and text and takes full responsibility for the publication.

\subsection*{Data availability}
This work is theoretical, and no data were generated or analysed, since all results follow from the derivations in the text and appendices.

\small
\setlength{\bibsep}{0pt}

\end{document}